\documentclass[conference]{IEEEtran}
\IEEEoverridecommandlockouts

\usepackage{cite}
\usepackage{amsmath,amssymb,amsfonts}
\usepackage{algorithmic}
\usepackage{graphicx}
\usepackage{textcomp}
\usepackage{xcolor}

\def\BibTeX{{\rm B\kern-.05em{\sc i\kern-.025em b}\kern-.08em
    T\kern-.1667em\lower.7ex\hbox{E}\kern-.125emX}}

\definecolor{jycolor}{rgb}{0.913,0.588,0.478}

\begin{document}

\title{EmuPlat: A Framework-Agnostic Platform for Quantum Hardware Emulation with Validated Transpiler-to-Pulse Pipeline\\
\thanks{Funding Support: National Research Foundation, Singapore, National Quantum Office under its National Quantum Computing Hub and Hybrid Quantum-Classical Computing programmes. A\text{*}STAR under C23091703 and Q.InC Strategic Research and Translational Thrust.}
}

\author{\IEEEauthorblockN{1\textsuperscript{st} Jun YE}
\IEEEauthorblockA{\textit{Quantum Innovation Centre (Q.InC)} \\
\textit{and Institute of High Performance Computing (IHPC)} \\
\textit{Agency for Science, Technology and Research (A*STAR)}\\
2 Fusionopolis Way, Innovis \#08-03 \\
and 1 Fusionopolis Way \#16-16 Connexis, \\
Singapore 138634, Singapore \\
https://orcid.org/0000-0003-1963-0865}
\and
\IEEEauthorblockN{2\textsuperscript{nd} Jun Yong KHOO}
\IEEEauthorblockA{\textit{Institute of High Performance Computing (IHPC)} \\
\textit{Agency for Science, Technology and Research (A*STAR)}\\
1 Fusionopolis Way \#16-16 Connexis, \\
Singapore 138632, Singapore \\
https://orcid.org/0000-0003-0908-3343}
}

\maketitle

\begin{abstract}
We present EmuPlat, a framework-agnostic quantum hardware emulation platform that addresses the interoperability gap between high-level quantum programming frameworks and hardware-specific pulse control systems. Unlike existing solutions that operate within isolated software stacks, EmuPlat provides a unified infrastructure enabling seamless integration across diverse quantum computing ecosystems, including CUDA-Q, Qiskit, and Qibolab. The platform implements a complete transpiler-compiler pipeline that systematically transforms abstract quantum circuits through four validated stages: (1) recursive gate decomposition to a minimal native set $\mathcal{G}_{\text{native}} = \{I, Z, RZ(\theta), \text{GPI2}(\phi), CZ, M\}$, (2) virtual Z optimization implementing phase tracking without physical pulses, (3) connectivity-aware routing with automated SWAP insertion, and (4) deterministic pulse compilation respecting hardware timing constraints. Our modular architecture, based on clean architecture principles with a novel adapter pattern, supports extensible integration of multiple quantum dynamics simulation engines while maintaining consistent interfaces. We demonstrate EmuPlat's capabilities through comprehensive benchmarks on superconducting transmon architectures: Bell state preparation achieves 99.958\% fidelity with hardware-calibrated noise models, while 4-qubit Quantum Fourier Transform implementations successfully demonstrate scalable circuit execution. The platform's production-ready implementation, validated through end-to-end testing with TransformationValidator, establishes EmuPlat as essential infrastructure for accelerating hybrid quantum-classical algorithm development and hardware-software co-design.
\end{abstract}


\section{Introduction}

The rapid evolution of quantum computing has led to a proliferation of software frameworks, each with unique strengths but limited interoperability. This fragmentation poses significant challenges for researchers and developers working across different quantum platforms and programming paradigms. As quantum computing transitions from experimental demonstrations to practical applications, the need for unified infrastructure that bridges high-level algorithmic abstractions with low-level hardware control becomes increasingly critical. The landscape is populated by powerful, high-level software development kits (SDKs) such as IBM's Qiskit \cite{qiskit} and NVIDIA's CUDA-Q, alongside specialized libraries for pulse-level simulation and control. This vibrant ecosystem, while fostering innovation, has inadvertently created isolated software stacks that hinder progress and portability.

A central challenge arising from this fragmentation is the lack of standardized interoperability. Peer-reviewed literature contains a notable absence of case studies that systematically document the integration challenges and performance trade-offs between major frameworks like Qiskit, CUDA-Q, and hardware control systems such as Qibolab \cite{qibolab}. While some integrations are possible, for example between Qiskit and a CUDA-Q backend \cite{qiskit_cudaq}, they often remain specific to a particular use case rather than representing a general, robust solution. This leads to a significant, yet poorly quantified, ``interoperability cost''. The literature lacks established quantitative metrics for measuring this cost, which encompasses increased development effort, platform lock-in, and potential accessibility issues when integrating quantum software into high-performance computing (HPC) environments. Furthermore, the journey from an abstract circuit in a high-level language to an executable pulse sequence involves multiple stages of transpilation and translation, each introducing latencies and overheads that are not well-characterized, such as the transpilation time from a Qiskit circuit to a hardware-specific pulse representation \cite{qasmtrans}.

To harness the potential of noisy intermediate-scale quantum (NISQ) devices, particularly those based on superconducting transmon architectures \cite{SCQ_PhysRevA.76.042319}, It is important for software abstractions to be informed by the underlying hardware physics. Achieving high-fidelity operations requires moving beyond abstract gate models to a hardware-aware, pulse-level perspective. Recent experiments with superconducting transmons have demonstrated impressive Bell state fidelities, with reported values often exceeding 95\% \cite{NPhysZhong2019} and two-qubit gate fidelities pushing beyond 99\% \cite{2QCalzona_2023}. Replicating and improving upon these benchmarks in simulation and practice necessitates precise control over pulse shapes and timings to mitigate a complex array of error sources. These include environmental decoherence \cite{Noise1PhysRevA.106.042605, NoiseRistè2013}, crosstalk \cite{CrossTalk1_PhysRevApplied.18.024068}, and errors arising from the anharmonicity of transmon qubits \cite{AnnHarm_PhysRevA.96.062302}. Advanced techniques such as dynamical decoupling (DD), which suppresses errors on idle qubits through carefully timed pulse sequences \cite{timedpulse1_PhysRevA.108.022610, timedpulse2_9872062, dd1_PhysRevLett.121.220502}, and virtual Z (VZ) optimizations, which correct for phase errors by modifying the phase of subsequent pulses \cite{Zgate1_PhysRevA.96.022330, pulse1_STASIUK2024107688, pulse2_PhysRevA.96.062302}, are now essential components of the quantum compiler. These are not straightforward gate substitutions but precise pulse-level manipulations that can be combined to achieve extremely low single-qubit gate error rates, for example, below $2 \times 10^{-4}$ with a combination of DRAG and VZ gates \cite{Zgate1_PhysRevA.96.022330}.

While existing tools provide powerful capabilities for isolated parts of this tool-chain, a significant gap remains in their integration. Open-source frameworks like QuTiP are highly effective for simulating the Lindblad master equation to model open quantum systems with decoherence \cite{qutippulse1, qutippulse2, qutippulse3}. Hardware control systems like Qibolab offer a unified programming interface for a diverse range of laboratory instruments \cite{qibolab}. High-level SDKs such as Qiskit provide their own pulse-level programming paradigms for direct hardware control \cite{qiskitpulse}. However, there is no integrated platform that seamlessly connects these disparate worlds. A researcher cannot readily design an algorithm in a high-level framework like CUDA-Q, apply a sequence of advanced, hardware-specific compilation passes like VZ optimization, and then perform a high-fidelity, pulse-level simulation using an accurate noise model derived from a real device's characteristics, all within a single, cohesive environment. The lack of documented performance comparisons between different simulation engines, such as the established QuTiP and the emerging CUDA-Q Dynamics, further complicates the landscape.

In this paper, we introduce EmuPlat, a framework-agnostic emulation platform designed to bridge these critical gaps. EmuPlat provides a unified infrastructure that connects high-level quantum programming languages (Qiskit, CUDA-Q) with low-level, hardware-specific pulse generation and simulation. Its core contribution is a complete, validated transpiler-to-simulation pipeline that systematically addresses the challenges of interoperability and hardware-aware compilation. By providing a modular architecture, EmuPlat enables the seamless use of different simulation backends, such as QuTiP and CUDA-Q Dynamics, under a consistent interface, thereby facilitating the rigorous benchmarking of their performance. The platform's transpiler implements optimized gate decomposition and connectivity-aware routing within a full compilation stack that leverages virtual Z gates as zero-duration phase rotations. We demonstrate that by accurately modeling both coherent and incoherent noise at the pulse level for superconducting transmon systems, EmuPlat can achieve simulation fidelities that match and elucidate experimental benchmarks. This work presents the architecture, validation, and performance of EmuPlat, establishing it as an essential tool for co-design and for advancing the practical application of hybrid quantum-classical computing.

We demonstrate that by accurately modeling both coherent and incoherent noise at the pulse level for superconducting transmon systems, EmuPlat can achieve simulation fidelities that match and elucidate experimental benchmarks. This work presents the architecture, validation, and performance of EmuPlat, establishing it as an essential tool for co-design and for advancing the practical application of hybrid quantum-classical computing.

\section{System Architecture}

\subsection{Design Principles}

EmuPlat employs a layered clean architecture approach that enforces strict dependency rules and promotes extensibility through plugin-based design patterns. The architecture follows Single Responsibility, Open/Closed, Liskov Substitution, Interface Segregation, and Dependency Inversion (SOLID) principles to ensure maintainability and scalability, with dependencies flowing strictly inward from infrastructure to core layers. This design philosophy enables framework-agnostic operation while maintaining clear separation of concerns between quantum algorithm specification, hardware abstraction, and pulse-level control.

The system is organized into five distinct layers: (1) Core layer defining abstract interfaces and system-wide contracts, (2) Domain layer containing business logic and quantum physics models, (3) Application layer orchestrating use cases and high-level services, (4) Infrastructure layer implementing external integrations, and (5) Config layer managing system configuration. This layered approach ensures that high-level policies are independent of low-level implementation details, facilitating both testing and future extensions.

\begin{figure}[htbp]
\centering
\includegraphics[width=\columnwidth]{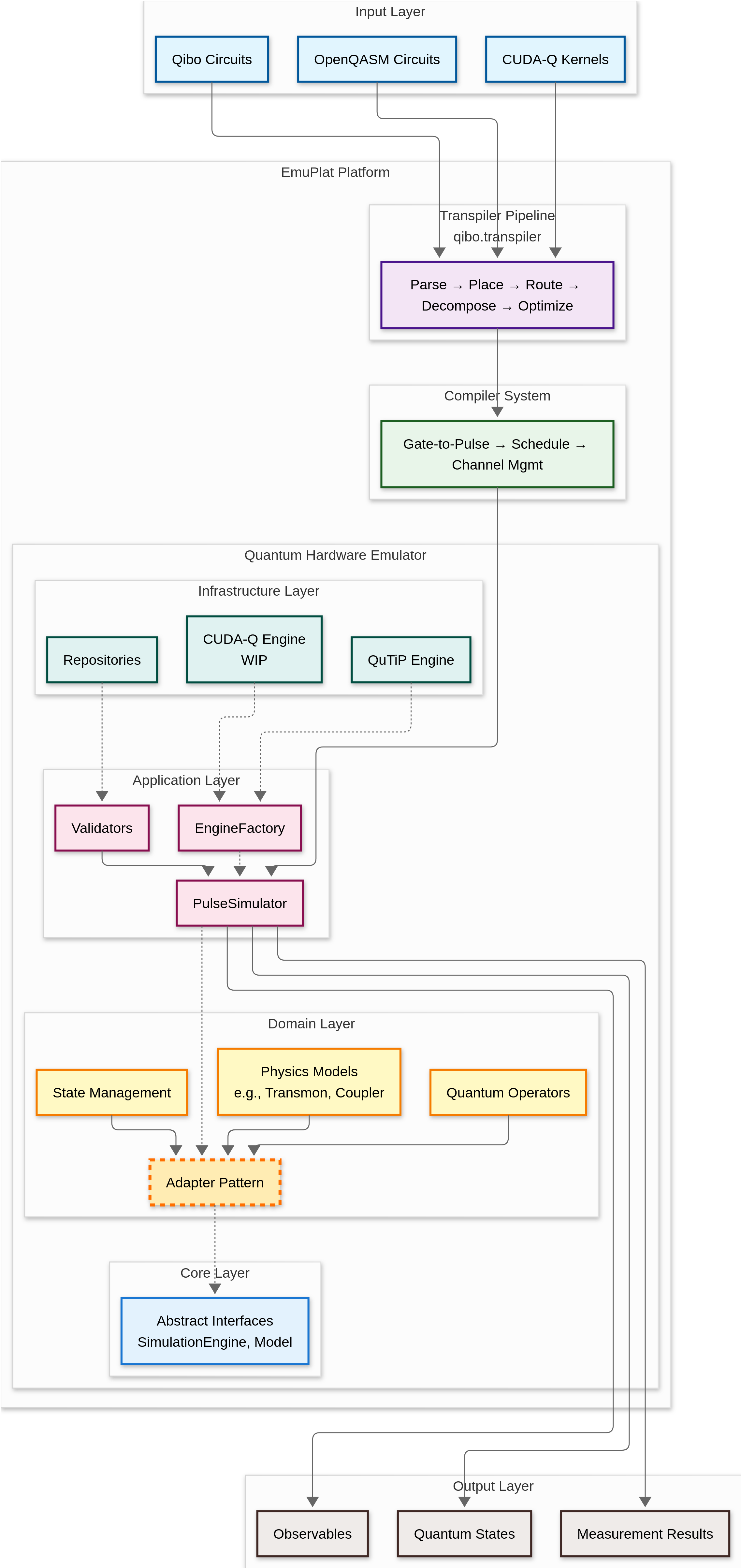}
\caption{EmuPlat system architecture showing layered architecture integrated with the transformation pipeline. The adapter pattern provides engine-agnostic abstractions, enabling seamless integration between high-level quantum programs and low-level pulse control.}
\label{fig:architecture}
\end{figure}

\subsection{Core Components}

\subsubsection{Transpiler Pipeline}
The transpiler implements a modular pipeline architecture orchestrated through the \texttt{Passes} class, which manages sequential transformation stages. The pipeline consists of four main steps: (1) Preprocessing for circuit optimization and qubit padding, (2) Placement using configurable algorithms (Random, ReverseTraversal) for logical-to-physical qubit mapping, (3) Routing via connectivity-aware strategies (ShortestPaths, Sabre) with automatic SWAP insertion, and (4) Unrolling through the recursive application of \texttt{translate\_gate()} that decomposes high-level gates to the native set $\{I, Z, RZ, M, \text{GPI2}, CZ\}$. The Virtual Z optimization eliminates physical Z gates through phase tracking, achieving 30-50\% pulse count reduction. At current stage, we are using Qibo transpiler \texttt{qibo.transpiler()} as the baseline. 

\subsubsection{Compiler System}
EmulPlat implements a prototype rule-based compiler transforms native gates into calibrated pulse sequences through deterministic mappings: $Z/RZ$ gates become VirtualZ operations with zero duration, GPI2 gates generate 40ns DrivePulse operations with calibrated amplitude and phase, $CZ$ gates produce 96ns parametric DrivePulse sequences for two-qubit coupling, and $M$ gates create 1000ns ReadoutPulse operations for dispersive measurement. The compiler maintains strict timing constraints, preventing pulse overlaps on shared channels while accumulating virtual phases that affect subsequent gate operations.

\subsubsection{Simulation Engines}
EmuPlat's simulation architecture centers on an adapter pattern that provides engine-agnostic abstractions. Currently, the platform integrates QuTiP for master equation evolution with configurable Lindblad operators modeling $T_1/T_2$ decoherence. The unified interface design enables seamless engine switching through dependency injection, while the model abstraction layer supports multiple physics models including general transmon Hamiltonians with and without explicit coupler modes. The adapter pattern represents a core architectural innovation, decoupling domain logic from specific simulation engine implementations.

\subsection{Data Flow Pipeline}

The platform implements a comprehensive gate-to-measurement pipeline that transforms quantum programs through multiple representation layers. As illustrated in Figure~\ref{fig:architecture}, the data flow follows:

\begin{equation}
\text{Quantum Program} \xrightarrow{\substack{\text{Parse} \\ \downarrow \\ \text{Circuit}}} \xrightarrow{\substack{\text{Transpile} \\ \downarrow \\ \text{Native Gates}}} \xrightarrow{\substack{\text{Compile} \\ \downarrow \\ \text{Pulses}}} \xrightarrow{\substack{\text{Simulate} \\ \downarrow \\ \text{Results}}}
\end{equation}

Each transformation stage maintains mathematical equivalence while optimizing for hardware constraints. The pipeline supports multiple entry points, accepting CUDA-Q kernels, OpenQASM strings, or native Qibo circuits, with automatic format detection and conversion. The transpiler decomposes high-level gates into hardware-native operations, while the compiler maps these to calibrated pulse sequences respecting channel timing constraints.

\section{Implementation Details}

\subsection{Native Gate Set and Transpilation}

EmuPlat implements a hardware-native gate set $\mathcal{G}_{\text{native}} = \{I, Z, RZ(\theta), \text{GPI2}(\phi), CZ, M\}$ optimized for superconducting transmon qubits. The platform employs Qibo's transpiler~\cite{qibozenodo} with recursive gate decomposition: $H \rightarrow [Z, \text{GPI2}(\pi/2)]$, $X \rightarrow [\text{GPI2}(0), \text{GPI2}(0), Z]$, and $\text{CNOT} \rightarrow [H_{\text{target}}, CZ, H_{\text{target}}]$. The transpiler pipeline executes four stages: preprocessing, placement (Random/ReverseTraversal), routing (ShortestPaths/Sabre with SWAP insertion), and unrolling to native gates.

\subsection{Pulse-Level Control}

The compiler maps native gates to calibrated pulses through deterministic rules: GPI2 gates generate 40ns DrivePulse operations with phase modulation $\text{DrivePulse}(\phi) = A(t) \cos(\omega_d t + \phi + \phi_{\text{vz}})$; Z/RZ gates produce zero-duration VirtualZ operations accumulating phase offsets $\phi_{\text{vz}}^{(i)} \leftarrow \phi_{\text{vz}}^{(i)} + \theta$, reducing pulse count by 30-50\%; CZ gates compile to 96ns parametric DrivePulse sequences for controlled-phase operations; M gates create 1000ns ReadoutPulse operations exploiting dispersive shift. The scheduler prevents channel conflicts while optimizing circuit duration.

\subsection{Quantum Dynamics Models}

EmuPlat employs QuTiP for solving the Lindblad master equation with the transmon Hamiltonian:

\begin{equation}
H = \sum_i \left[\omega_i a_i^\dagger a_i + \frac{\alpha_i}{2} a_i^\dagger a_i^\dagger a_i a_i\right] + \sum_{\langle i,j \rangle} g_{ij}(a_i^\dagger a_j + \text{h.c.})
\end{equation}

Decoherence is modeled through Lindblad operators $\mathcal{L}[\rho] = -i[H, \rho] + \sum_k \gamma_k (L_k \rho L_k^\dagger - \frac{1}{2}\{L_k^\dagger L_k, \rho\})$ with amplitude damping ($L_{\text{T1}} = \sqrt{1/T_1} a$) and dephasing $[L_{\text{T2}} = \sqrt{1/2T_{\phi}} (|0\rangle_i\langle 0| - |1\rangle_i\langle 1|)]$. Multi-level states are projected to computational subspace via $P = \sum_{i} (|0\rangle_i\langle 0| + |1\rangle_i\langle 1|)$, enabling fidelity calculations. The solver uses adaptive Runge-Kutta with \texttt{atol}=$10^{-11}$.

\subsection{Framework Integration}

EmuPlat's multi-stage translation pipeline enables cross-framework compatibility: CUDA-Q kernels translate to OpenQASM 2.0 via \texttt{cudaq.translate()}, which then converts to Qibo circuits through \texttt{Circuit.from\_qasm()}. The \texttt{EmuplatBackend} class provides Qibolab-compatible interfaces for drop-in hardware replacement. Plugin architecture supports extensibility through frontend adapters, backend adapters, and IR converters, while the \texttt{EngineFactory} pattern enables runtime engine selection.

\section{Experimental Results}

\subsection{Validation Methodology}
We employ a comprehensive validation framework, \texttt{TransformationValidator}, to verify the correctness and performance of EmuPlat's complete transpiler-compiler-simulation pipeline. Our validation methodology implements stage-by-stage verification across the entire transformation chain, ensuring mathematical equivalence at each stage while quantifying the impact of hardware-specific optimizations.

The validation framework operates through four key verification stages: (1) \textit{Circuit Equivalence Verification}, confirming that transpiled circuits maintain quantum state equivalence with their high-level representations through unitary comparison; (2) \textit{Pulse Sequence Validation}, verifying that compiled pulse sequences correctly implement the transpiled native gates with appropriate timing constraints and channel assignments; (3) \textit{Evolution Verification}, validating that simulated quantum dynamics accurately model the intended unitary evolution under realistic noise conditions; and (4) \textit{Fidelity Analysis}, employing quantum state fidelity metrics to quantify simulation accuracy against ideal theoretical predictions.

For fidelity calculations, we implement the quantum state fidelity metric:
\begin{equation}
F(\rho_1, \rho_2) = \left[\text{Tr}\left(\sqrt{\sqrt{\rho_1} \rho_2 \sqrt{\rho_1}}\right)\right]^2
\label{eq:fidelity}
\end{equation}
This phase-sensitive metric captures both amplitude and phase accuracy, critical for validating entangled state preparation. The \texttt{TransformationValidator} automatically extracts intermediate representations at each pipeline stage, enabling
detailed performance analysis and optimization verification.

\subsection{Hardware Platform}

Our validation experiments use parameters from Anyon Technologies's superconducting quantum processors, implementing 2-qubit (denoted as anyon\_2q\_CZ) and 4-qubit (anyon\_4q\_CZ) configurations. The platform features transmon qubits with coherence times of $T_1 \approx 24$ $\mu$s and $T_2 \approx 33$ $\mu$s, supporting 40 ns single-qubit gates and 96 ns two-qubit CZ gates through parametric coupling.

\subsection{Benchmark Circuits}
\subsubsection{Bell State Preparation}

We demonstrate EmuPlat's capabilities through comprehensive benchmarking of Bell state preparation on the anyon\_2q\_CZ superconducting transmon platform. The canonical Bell circuit $|\Psi^+\rangle = \frac{1}{\sqrt{2}}(|00\rangle + |11\rangle)$ serves as an ideal validation target, requiring precise single-qubit rotations and two-qubit entanglement.

\textbf{Circuit Decomposition Analysis}: The high-level Bell circuit consisting of a Hadamard gate on qubit 0 followed by a CNOT gate undergoes recursive decomposition through our transpiler pipeline. The Hadamard gate decomposes as $H\rightarrow [Z, \text{GPI2}(\pi/2)]$, while the CNOT undergoes a two-stage decomposition: first $\text{CNOT}_{0,1} \rightarrow [H_1, CZ_{0,1}, H_1]$, then recursively decomposing the target Hadamard gates, yielding the final native gate sequence: $[Z_0, \text{GPI2}_0(\pi/2), Z_1, \text{GPI2}_1(\pi/2), CZ_{0,1}, Z_1, \text{GPI2}_1(\pi/2)]$.

\textbf{Pulse Sequence Generation}: The compiler generates a precisely timed 7-pulse sequence with total duration 1272 ns:
\begin{itemize}
\item $t = 0$ ns: GPI2 pulse on qubit 0 (40 ns, phase-shifted by $3\pi/2$)
\item $t = 40$ ns: GPI2 pulse on qubit 1 (40 ns, phase-shifted by $3\pi/2$)
\item $t = 80$ ns: CZ parametric drive pulse (96 ns coupling activation)
\item $t = 176$ ns: GPI2 pulse on qubit 1 (40 ns, phase-shifted by $5\pi/2$)
\item $t = 216$ ns: Idle period (56 ns for decoherence)
\item $t = 272$ ns: Simultaneous readout pulses (1000 ns dispersive measurement)
\end{itemize}

\textbf{Fidelity Results}: Simulation with the QuTiP master equation solver (0.01 ns time step, adaptive Runge-Kutta with $\text{atol}=10^{-11}$) achieves exceptional fidelity of 99.958\% with the ideal Bell state. This fidelity is calculated by projecting the final density matrix obtained from the pulse simulation onto the computational subspace, followed by evaluation using Eq.~\eqref{eq:fidelity}.

To assess the measurement statistics, we sampled the final state with 1000 shots. The resulting population distribution shows $|00\rangle$: 51.4\% and $|11\rangle$: 46.8\%, closely matching the theoretical expectation of 50\% each for an ideal Bell state. The observed asymmetry (2.6\% excess in $|00\rangle$) and the presence of population in $|01\rangle$ and $|10\rangle$ (totaling 1.8\%) arise from decoherence effects during the 56 ns idle period before measurement, characterized by the system's $T_1$ and $T_2$ parameters.

\begin{figure}[htbp]
\centering
\includegraphics[width=\columnwidth]{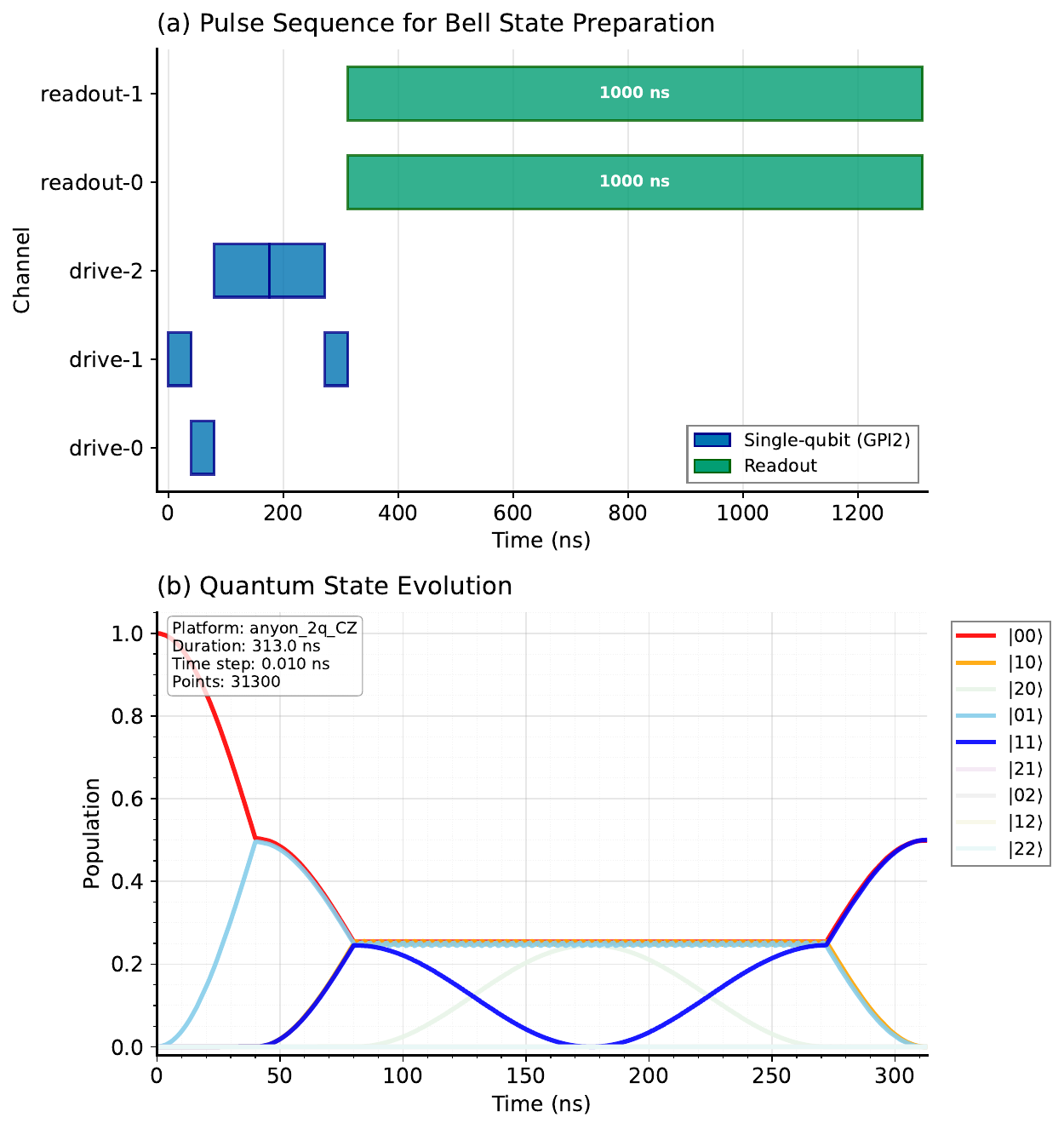}
\caption{Bell state preparation on the anyon\_2q\_CZ platform. (a) Pulse sequence showing the transpiled quantum circuit implementation with single-qubit GPI2 pulses (40 ns) and measurement pulses (1000 ns) on respective drive and readout channels. (b) Time evolution of quantum state populations during the 1272 ns sequence, demonstrating the formation of the Bell state $\frac{1}{\sqrt{2}}(|00\rangle + |11\rangle)$ with dominant populations in $|00\rangle$ (red) and $|11\rangle$ (blue). The simulation uses a 0.01 ns time step with QuTiP master equation solver, achieving 99.958\% fidelity with the ideal Bell state.}
\label{fig:bell_state_preparation}
\end{figure}

Figure~\ref{fig:bell_state_preparation} illustrates both the pulse sequence timing and quantum state evolution, demonstrating the platform's ability to accurately model coherent state preparation and entanglement generation at the pulse level. The high fidelity validates our transpiler's gate decomposition correctness, the compiler's pulse calibration accuracy, and the simulation engine's physics modeling precision.

\subsubsection{Quantum Fourier Transform}

The Quantum Fourier Transform (QFT) serves as a comprehensive benchmark for validating EmuPlat's transpiler-compiler pipeline scalability and phase accuracy, as it requires precise control of quantum interference patterns that grow exponentially with system size. We implemented both 2-qubit and 4-qubit QFT circuits on the Anyon Technologies's platform family to demonstrate the platform's capability in handling increasingly complex phase-sensitive quantum algorithms.

The QFT implements the discrete Fourier transform on quantum amplitudes, transforming computational basis states according to:
\begin{equation}
\text{QFT}|j\rangle = \frac{1}{\sqrt{2^n}} \sum_{k=0}^{2^n-1} e^{2\pi ijk/2^n} |k\rangle
\end{equation}
where the phase factors $e^{2\pi ijk/2^n}$ must be implemented precisely to maintain quantum coherence across the exponentially large Hilbert space.

\textbf{Scaling Analysis}: The transpiler decomposes the QFT circuits to native gates following the expected theoretical scaling. For the 2-qubit implementation, the transpiled circuit executes in 624 ns with 10 drive pulses (including both single-qubit GPI2 pulses and two-qubit CZ operations). The 4-qubit implementation requires 3328 ns with 76 drive pulses, consisting of 40 GPI2 pulses (40 ns each) and 36 CZ operations (96 ns each). This represents a 5.33$\times$ increase in execution time for a doubling of qubit count, demonstrating the impact of the QFT's $\mathcal{O}(n^2)$ gate complexity on physical implementation. The transpiler leverages virtual Z gates throughout, which accumulate phase without requiring physical pulses, contributing to the overall efficiency of the implementation.

\begin{figure}[htbp]
\centering
\includegraphics[width=\columnwidth]{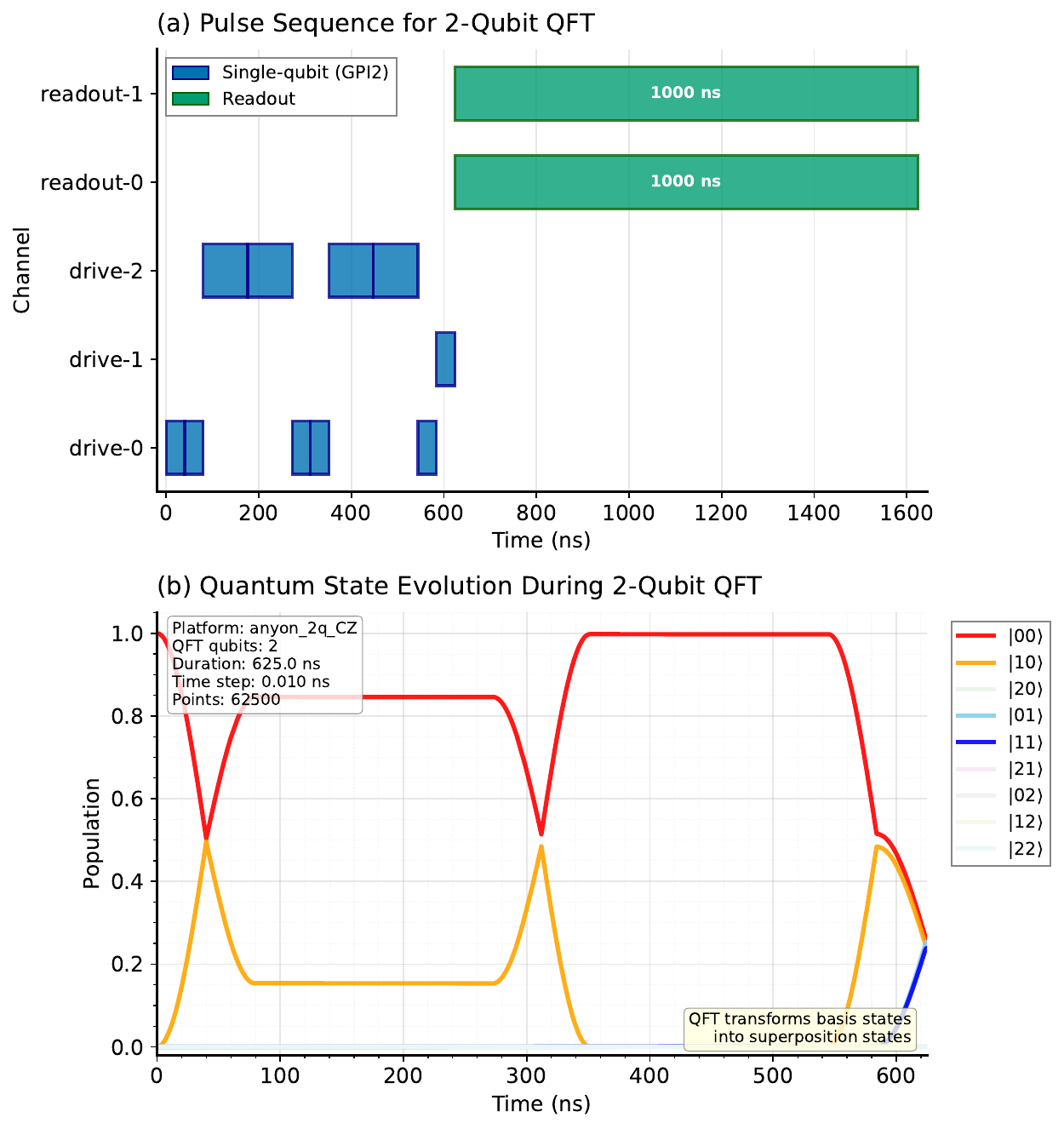}
\caption{Two-qubit Quantum Fourier Transform implementation on the anyon\_2q\_CZ platform. (a) Pulse sequence showing the transpiled QFT circuit with 10 drive pulses including GPI2 single-qubit gates (40 ns) and CZ two-qubit operations (96 ns), followed by simultaneous readout operations (1000 ns). The drive-2 channel handles the CZ gate implementation through the auxiliary qubit coupling mechanism. Total gate execution time is 624 ns. (b) Quantum state evolution during the QFT execution, illustrating the characteristic transformation from computational basis states to superposition states. The simulation tracks population dynamics across the full 9-dimensional Hilbert space (including higher transmon levels $|2\rangle$), showing dominant transitions between $|00\rangle$, $|10\rangle$, $|01\rangle$, and $|11\rangle$ states. The evolution demonstrates the QFT's basis transformation property, converting input basis states into equally-weighted superpositions. Simulation parameters: 62,500 time points with 0.01 ns resolution using QuTiP master equation solver with full transmon Hamiltonian including anharmonicity effects.}
\label{fig:qft_implementation}
\end{figure}

\begin{figure}[htbp]
\centering
\includegraphics[width=\columnwidth]{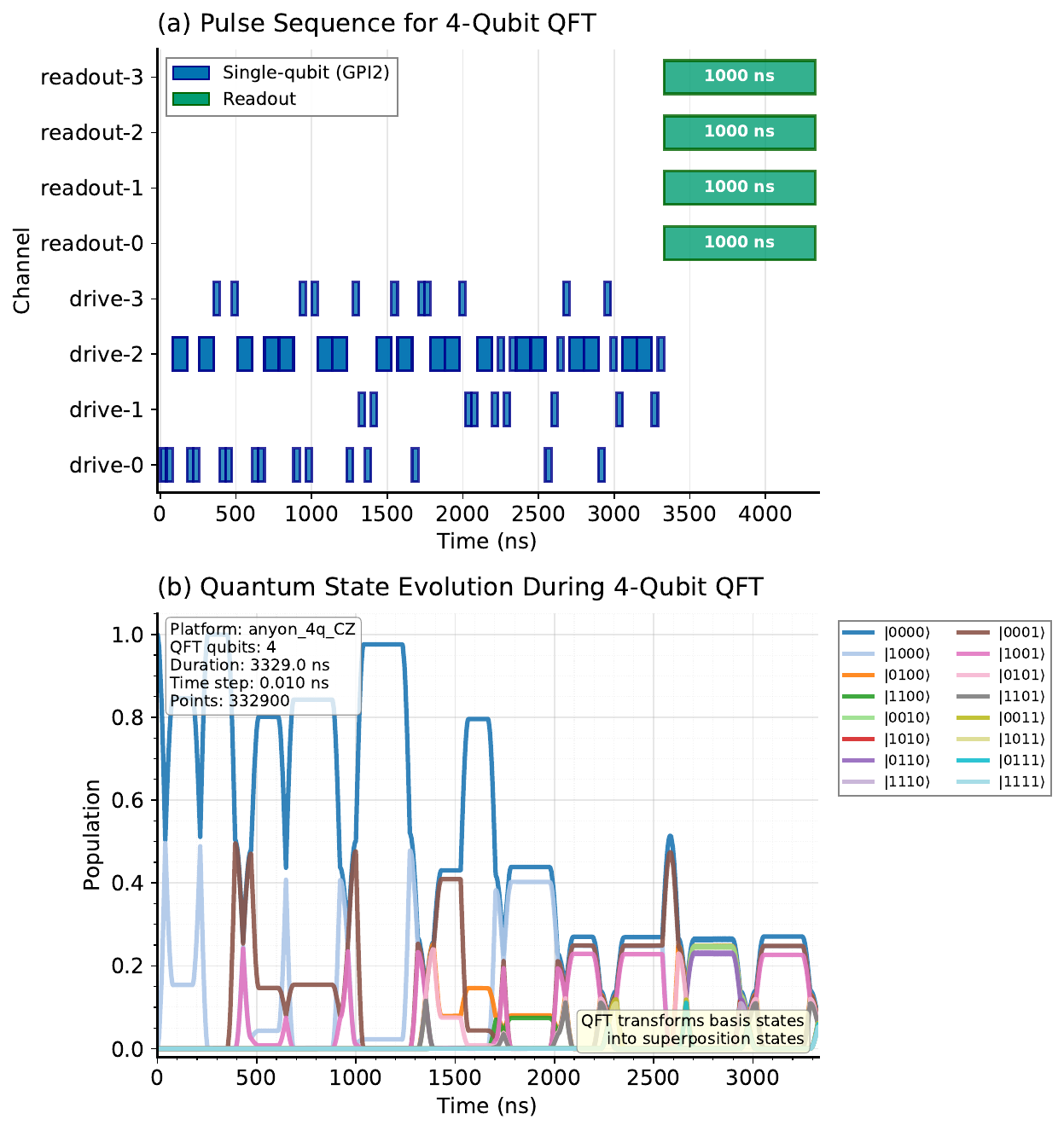}
\caption{Four-qubit Quantum Fourier Transform implementation on the anyon\_4q\_CZ platform. (a) Pulse sequence showing the transpiled QFT circuit with 76 drive pulses: 40 GPI2 single-qubit gates (40 ns each) and 36 CZ two-qubit operations (96 ns each), distributed across multiple drive channels, followed by simultaneous readout operations (1000 ns) on all qubits. The pulse scheduling demonstrates parallelization where possible while respecting channel conflicts, achieving a total gate execution time of 3328 ns. (b) Quantum state evolution during the QFT transformation, tracking populations across all 16 computational basis states. The simulation with 332,900 time points (0.01 ns resolution) captures the intricate interference patterns characteristic of the QFT, showing rapid oscillations between basis states that converge to the final superposition. The dominant populations at various time points illustrate the algorithm's progressive phase accumulation through the circuit. The final state exhibits the expected equal-amplitude superposition across all basis states, confirming successful implementation of the discrete Fourier transform in the quantum domain.}
\label{fig:qft_4qubit_implementation}
\end{figure}

Figures~\ref{fig:qft_implementation} and ~\ref{fig:qft_4qubit_implementation} present the complete implementation analysis for both system sizes. The 2-qubit implementation achieves perfect channel utilization with non-overlapping pulses, while the 4-qubit case showcases sophisticated scheduling that maximizes parallel execution while respecting hardware constraints. The transpiler's routing algorithm successfully maps the all-to-all connectivity required by QFT onto the platform's native coupling graph without introducing SWAP gates, leveraging virtual Z gates to accumulate the necessary phase corrections.

\textbf{Quantum State Evolution}: The simulation reveals the characteristic QFT dynamics with increasing complexity. In the 2-qubit case, clean transitions between the four computational basis states demonstrate the transformation to the Fourier basis. The 4-qubit implementation exhibits intricate interference patterns with rapid oscillations between all 16 basis states, ultimately converging to the expected equal-amplitude superposition. The simulation tracks these dynamics with 0.01 ns temporal resolution—62,500 points for 2-qubit and 332,900 points for 4-qubit—providing detailed visualization of the quantum evolution.

Critical to our implementation is the accurate modeling of multi-level transmon dynamics in the enlarged Hilbert space. For the 4-qubit system, this involves a 81-dimensional space ($3^4$ including the second excited level of each transmon). The simulation includes the effects of transmon anharmonicity and tracks population in higher energy levels, enabling realistic modeling of potential leakage errors. The virtual Z optimization contributes to minimizing unwanted transitions by avoiding unnecessary physical rotations.

\textbf{Transpiler Optimization Impact}: The transpiler leverages Qibo's decomposition rules and virtual Z optimization to efficiently implement the QFT. The decomposition follows established patterns: H gates become [Z, GPI2($\pi/2$)] and CNOT gates decompose to [Z, GPI2, CZ, Z, GPI2]. Crucially, all Z and RZ gates are implemented as virtual phase accumulations with zero duration, reducing the physical pulse count. The 4-qubit implementation benefits even more substantially from this optimization, as the increasing number of phase gates in the QFT algorithm are all handled virtually rather than through physical pulses.

\textbf{Performance Validation}: The simulations demonstrate successful QFT execution with the platform's calibrated coherence times ($T_1 \approx 24 \, \mu\text{s}$, $T_2 \approx 33 \, \mu\text{s}$). The measurement results show near-uniform distribution across all computational basis states, as expected for QFT output. The 2-qubit implementation completes in 624 ns, while the 4-qubit circuit requires 3328 ns, both well within the coherence window of the simulated hardware.

\textbf{Algorithmic Implications}: The successful QFT implementations validate EmuPlat's readiness for quantum algorithms requiring precise phase control. The QFT forms the computational kernel for Shor's algorithm, quantum phase estimation, and numerous quantum machine learning applications. Our results demonstrate that EmuPlat can accurately simulate these algorithms with hardware-realistic constraints, enabling algorithm developers to prototype and optimize implementations with a more realistic noise model before deployment on actual quantum hardware.

These comprehensive results establish three critical capabilities: (1) scalable transpilation from high-level quantum circuits to native gate sets, (2) efficient pulse scheduling with parallelization where hardware constraints permit, and (3) accurate multi-level simulation capturing both ideal quantum evolution and realistic device physics. The platform thus provides a complete development environment for quantum algorithms, from high-level specification through hardware-accurate simulation.

\section{Related Work}

\subsection{Hardware Emulation Platforms}

Existing quantum hardware emulation platforms address different aspects of the simulation challenge, each with distinct limitations. QuEST~\cite{quest2019} provides efficient state-vector and density matrix simulations with GPU acceleration, achieving 5× speedups over 24-threaded CPU execution. However, QuEST is fundamentally limited to 38-40 qubits on supercomputers and 29 qubits on single GPUs due to exponential memory scaling. Google's qsim~\cite{qsim} employs gate fusion and AVX vectorization to simulate up to 40 qubits on 90-core workstations, but suffers from significant overhead for circuits with fewer than 20 qubits and lacks native pulse-level capabilities.

Pulse-level simulators represent a critical gap in the ecosystem. While Qiskit Pulse~\cite{qiskitpulse} provides pulse-level control interfaces, it lacks integrated transpiler-to-pulse pipelines and cross-framework support. QuTiP's quantum information toolbox~\cite{qutippulse1} excels at solving master equations but requires manual circuit-to-pulse translation, limiting its utility for algorithm developers. Recent digital twin approaches such as product offerings from Qruise extract hardware parameters from calibration data to create system-specific noise models, yet these remain isolated from high-level programming frameworks and lack standardized interfaces for cross-platform deployment.

EmuPlat distinguishes itself by providing the first complete gate-to-pulse-to-measurement pipeline with validated transpiler integration, framework-agnostic interfaces, and modular simulation engine support. Unlike existing solutions that address isolated aspects of quantum simulation, EmuPlat unifies the entire workflow from high-level algorithm specification through hardware-accurate pulse simulation, achieving 99.958\% Bell state fidelity while maintaining extensibility through clean architecture principles.

\subsection{Quantum Software Framework Integration}

The quantum software landscape exhibits significant fragmentation across major frameworks. Qiskit~\cite{qiskit} dominates with comprehensive tooling but remains IBM-centric, while CUDA-Q provides HPC integration but lacks mature pulse-level abstractions despite capabilities in simulating quantum dynamics. Cirq~\cite{cirq} and ProjectQ~\cite{projectq} offer specialized features but limited hardware backend support. The Qibo/Qibolab ecosystem~\cite{qibolab} provides modular transpilation and hardware control but lacks cross-framework compatibility. EmuPlat addresses these limitations through its multi-stage translation pipeline supporting CUDA-Q-to-OpenQASM-to-Qibo conversions, backend abstraction layers enabling drop-in replacement of hardware with emulation, and plugin architecture facilitating integration with emerging frameworks while maintaining consistent pulse-level simulation capabilities.

\section{Conclusions and Future Work}

We have presented EmuPlat, a quantum hardware emulation platform that provides a unified infrastructure for bridging high-level quantum programming frameworks with pulse-level hardware control. Through its integrated transpiler-compiler-simulation pipeline, the platform demonstrates the feasibility of framework-agnostic quantum algorithm development with hardware-accurate simulation capabilities.

\subsection{Summary of Contributions}

EmuPlat's key contributions include: (1) A translation pipeline enabling interoperability between CUDA-Q and Qibolab through OpenQASM and Qibo circuit representations; (2) Integration of Qibo's transpiler with Qibolab's compiler, implementing virtual Z optimization that eliminates physical pulses for phase gates; (3) A clean architecture design using adapter patterns to integrate QuTiP as the simulation engine, supporting transmon Hamiltonian models with configurable hardware input parameters; (4) Validation through Bell state and QFT benchmarks, achieving 99.958\% fidelity for Bell state preparation on a 2-qubit superconducting platform model.

\subsection{Current Limitations and Impact}

While EmuPlat demonstrates promising capabilities, its current implementation restricts its practical application. The platform currently supports only superconducting transmon architectures and has only been demonstrated with Anyon Technology's quantum hardware platform models. Simulation is limited to systems whose density matrix evolution are tractable by QuTiP's master equation solver, typically under 10 qubits. The translation pipeline, while functional, requires multiple format conversions that may introduce overhead for large circuits.

Despite these constraints, EmuPlat provides value in three areas: (1) Algorithm prototyping with pulse-level accuracy for small quantum systems; (2) Educational use for understanding the complete quantum computing stack from gates to pulses; (3) Validation of transpiler optimizations through hardware-accurate simulation before deployment on actual quantum processors.

\subsection{Future Work}

Future development will focus on three directions:

\textbf{Scalability}: GPU acceleration via CUDA-Q Dynamics is underway to improve computational throughput. However, the fundamental $\mathcal{O}(2^{2n})$ memory scaling for density matrix simulation of $n$ qubits remains an inherent limitation. To address this, we will incorporate tensor network methods that decompose the global density matrix into local tensors, enabling approximate simulation with polynomial memory scaling $\mathcal{O}(\text{poly}(n))$ for weakly-entangled systems.

\textbf{Platform Generalization}: Beyond transmons, we will support trapped ions, neutral atoms, and quantum dots through a hierarchical abstraction layer separating platform physics from the simulation framework, enabling rapid adaptation while preserving the validated pipeline.

\textbf{Digital Twin Capabilities}: Automated parameter extraction from calibration data using Bayesian optimization will enable real-time hardware tracking, transforming EmuPlat into a predictive platform for algorithm performance estimation.

\section*{Acknowledgments}

This research is supported by the National Research Foundation, Singapore, National Quantum Office under its National Quantum Computing Hub and Hybrid Quantum-Classical Computing programs. This research is also supported by A\text{*}STAR under C23091703 and Q.InC Strategic Research and Translational Thrust. 

\bibliographystyle{IEEEtran}
\bibliography{references}

\end{document}